\documentclass[twoside,twocolumn,9pt]{article}
\usepackage{extsizes}
\usepackage[super,sort&compress,comma]{natbib} 
\usepackage[version=3]{mhchem}
\usepackage[left=1.5cm, right=1.5cm, top=1.785cm, bottom=2.0cm]{geometry}
\usepackage{balance}
\usepackage{mathptmx}
\usepackage{sectsty}
\usepackage{graphicx} 
\usepackage{lastpage}
\usepackage[format=plain,justification=justified,singlelinecheck=false,font={stretch=1.125,small,sf},labelfont=bf,labelsep=space]{caption}
\usepackage{float}
\usepackage{fancyhdr}
\usepackage{fnpos}
\usepackage[english]{babel}
\addto{\captionsenglish}{%
  
}
\usepackage{array}
\usepackage{droidsans}
\usepackage{charter}
\usepackage[T1]{fontenc}
\usepackage[usenames,dvipsnames]{xcolor}
\usepackage{setspace}
\usepackage[compact]{titlesec}
\usepackage{hyperref}

\usepackage{epstopdf}

\usepackage{physics}
\usepackage{notes2bib}
\usepackage{bbold}

\definecolor{cream}{RGB}{222,217,201}
\definecolor{blue}{rgb}{0.01, 0.31, 0.59}

\begin{document}

\pagestyle{fancy}
\thispagestyle{plain}
\fancypagestyle{plain}{
\renewcommand{\headrulewidth}{0pt}
}

\makeFNbottom
\makeatletter
\renewcommand\LARGE{\@setfontsize\LARGE{15pt}{17}}
\renewcommand\Large{\@setfontsize\Large{12pt}{14}}
\renewcommand\large{\@setfontsize\large{10pt}{12}}
\renewcommand\footnotesize{\@setfontsize\footnotesize{7pt}{10}}
\makeatother

\renewcommand{\thefootnote}{\fnsymbol{footnote}}
\renewcommand\footnoterule{\vspace*{1pt}%
\color{cream}\hrule width 3.5in height 0.4pt \color{black}\vspace*{5pt}} 
\setcounter{secnumdepth}{5}

\makeatletter 
\renewcommand\@biblabel[1]{#1}            
\renewcommand\@makefntext[1]%
{\noindent\makebox[0pt][r]{\@thefnmark\,}#1}
\makeatother 
\renewcommand{\figurename}{\small{Fig.}~}
\sectionfont{\sffamily\Large}
\subsectionfont{\normalsize}
\subsubsectionfont{\bf}
\setstretch{1.125} 
\setlength{\skip\footins}{0.8cm}
\setlength{\footnotesep}{0.25cm}
\setlength{\jot}{10pt}
\titlespacing*{\section}{0pt}{4pt}{4pt}
\titlespacing*{\subsection}{0pt}{15pt}{1pt}

\fancyfoot{}
\fancyfoot[LO,RE]{\vspace{-7.1pt}\includegraphics[height=9pt]{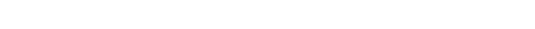}}
\fancyfoot[CO]{\vspace{-7.1pt}\hspace{13.2cm}\includegraphics{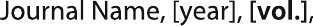}}
\fancyfoot[CE]{\vspace{-7.2pt}\hspace{-14.2cm}\includegraphics{head_foot/RF}}
\fancyfoot[RO]{\footnotesize{\sffamily{1--\pageref{LastPage} ~\textbar  \hspace{2pt}\thepage}}}
\fancyfoot[LE]{\footnotesize{\sffamily{\thepage~\textbar\hspace{3.45cm} 1--\pageref{LastPage}}}}
\fancyhead{}
\renewcommand{\headrulewidth}{0pt} 
\renewcommand{\footrulewidth}{0pt}
\setlength{\arrayrulewidth}{1pt}
\setlength{\columnsep}{6.5mm}
\setlength\bibsep{1pt}

\makeatletter 
\newlength{\figrulesep} 
\setlength{\figrulesep}{0.5\textfloatsep} 

\newcommand{\topfigrule}{\vspace*{-1pt}%
\noindent{\color{cream}\rule[-\figrulesep]{\columnwidth}{1.5pt}} }

\newcommand{\botfigrule}{\vspace*{-2pt}%
\noindent{\color{cream}\rule[\figrulesep]{\columnwidth}{1.5pt}} }

\newcommand{\dblfigrule}{\vspace*{-1pt}%
\noindent{\color{cream}\rule[-\figrulesep]{\textwidth}{1.5pt}} }

\makeatother

\twocolumn[
  \begin{@twocolumnfalse}
{
\hfill\raisebox{0pt}[0pt][0pt]{\includegraphics[height=55pt]{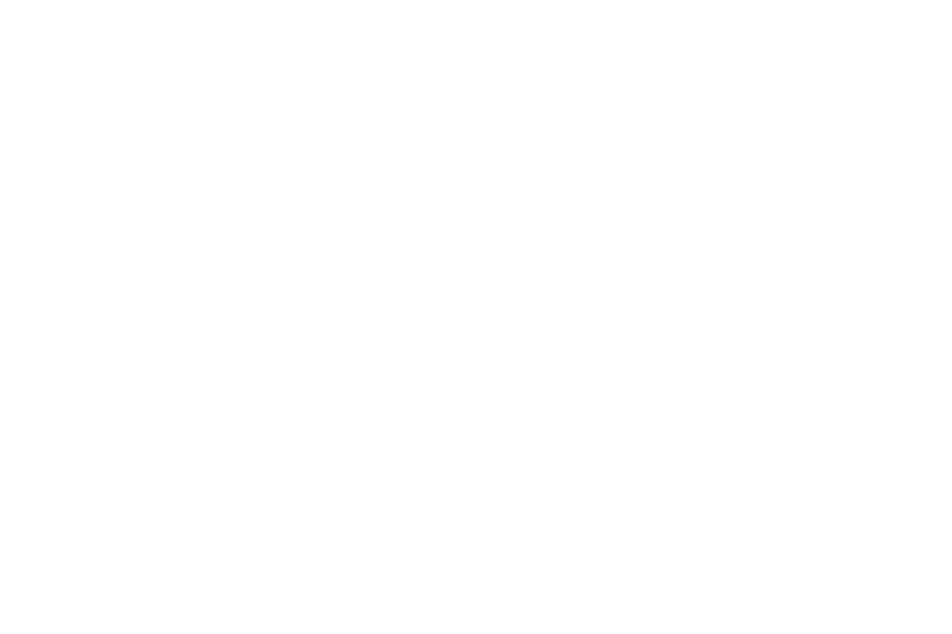}}\\[1ex]
\includegraphics[width=18.5cm]{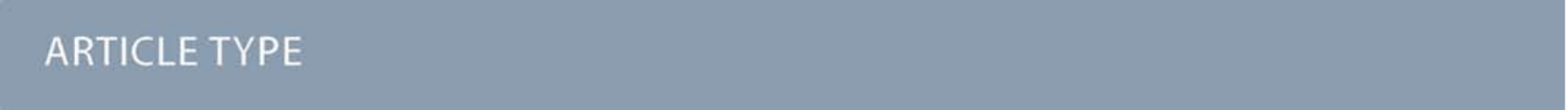}}\par
\vspace{1em}
\sffamily
\begin{tabular}{m{4.5cm} p{13.5cm} }

\includegraphics{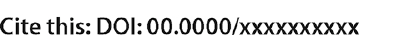} & 
\noindent\LARGE{\textbf{\emph{In silico} study of liquid crystalline phases formed by bent-shaped molecules with excluded-volume type interactions}} \\
\vspace{0.3cm} & \vspace{0.3cm} \\

 & \noindent\large{Piotr Kubala,\textit{$^{a}$} Wojciech Tomczyk,$^{\ast}$\textit{$^{ab}$} and Micha\l{} Cie\'sla\textit{$^{a}$}} \\

\includegraphics{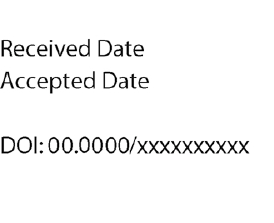} & \noindent\normalsize{We have numerically studied a liquid composed of achiral, bent-shaped molecules built of tangent spheres. The system is known to spontaneously break mirror symmetry, as it forms a macroscopically chiral, twist-bend nematic phase [Phys. Rev. Lett. {\bf{115}}, 147801 (2015)]. Here, we have examined full phase diagram of such liquid and observed several phases characterized by orientational and/or translational ordering of molecules. Apart from conventional nematic, smectic A, and the above-mentioned twist-bend nematic phase, we have identified antiferroelectric smectic A phase. For large densities and high degree of molecule's structural bend, another smectic phase emerged, where the polarization vector rotates within a single smectic layer. These results were confirmed using both Monte Carlo and molecular dynamics simulations.} \\

\end{tabular}

 \end{@twocolumnfalse} \vspace{0.6cm}

  ]

\renewcommand*\rmdefault{bch}\normalfont\upshape
\rmfamily
\section*{}
\vspace{-1cm}


\footnotetext{\textit{$^{a}$ Institute of Theoretical Physics, Jagiellonian University, \L{}ojasiewicza 11, 30-348 Krak\'{o}w, Poland}}
\footnotetext{\textit{$^{b}$ Jerzy Haber Institute of Catalysis and Surface Chemistry, Polish Academy of Sciences, Niezapominajek 8, 30-239 Krak\'{o}w, Poland. E-mail: wojciech.tomczyk@ikifp.edu.pl }}



\section{Introduction}

\begin{figure}[h!]
    \centering
    \includegraphics[width=.9\columnwidth]{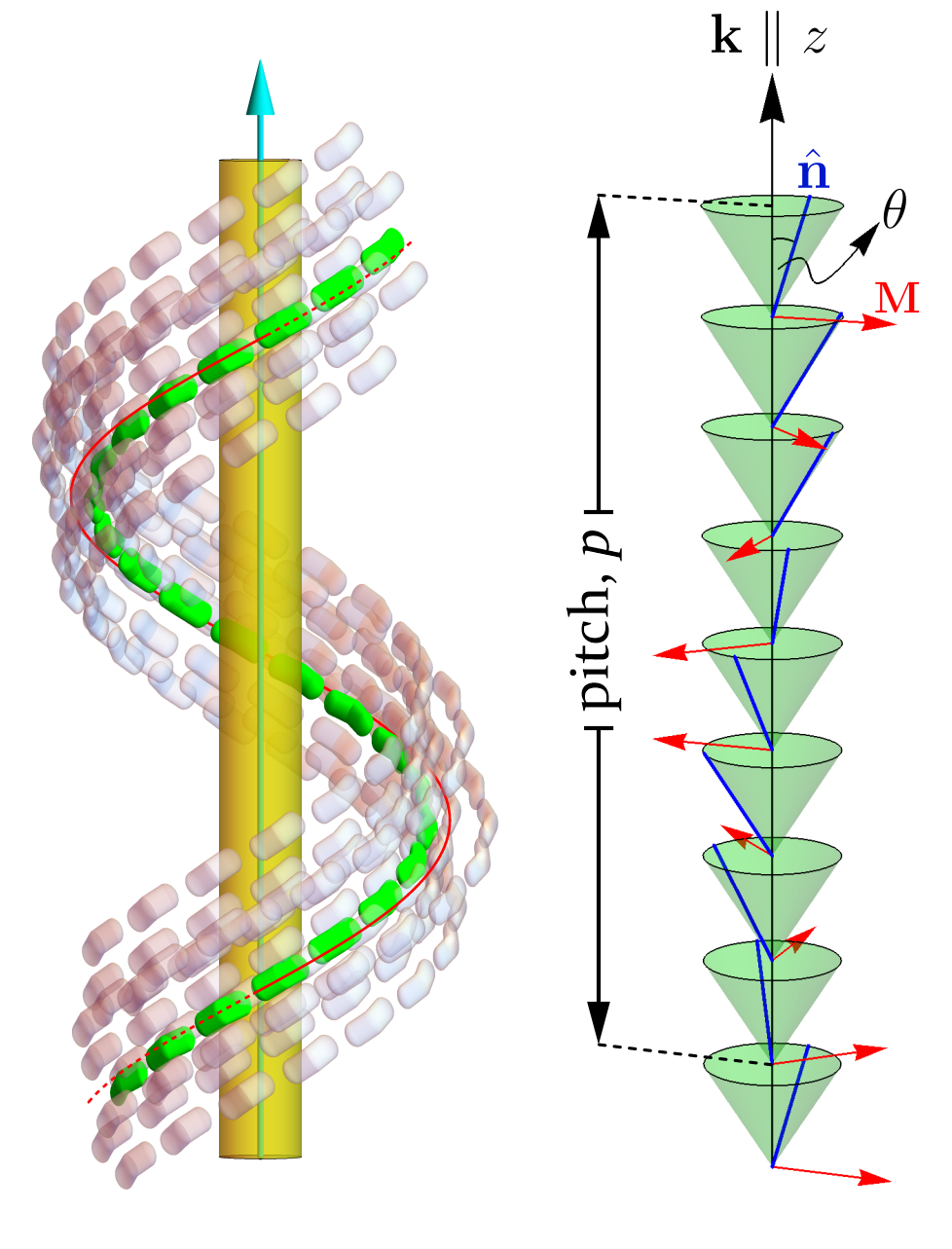}
\caption{Schematic depiction of the twist-bend nematic phase characterized by the finite pitch $p$ and heliconical tilt angle $\theta$ between the director $\vu{n}$ and the wavevector $\mathbf{k}$-axis. The polarization $\mathbf{M}$ is perpendicular to both $\vu{n}$ and $\mathbf{k}$, and is precessing around $\mathbf{k}$ along with $\vu{n}$. (Left) Reproduced from Ref.~\cite{Tomczyk2020} with permission from the Royal Society of Chemistry. (Right) Adapted from Ref.~\cite{Longa2020} published under ACS AuthorChoice with CC-BY license.}
    \label{fig:1}
\end{figure}

Comprehension of spontaneous mirror symmetry breaking (SMSB) phenomenon is essential for understanding fundamental aspects such as life itself \cite{Fujii2004, Viedma2007, Ruiz-Mirazo2014}. Investigations revolving around the origins and corollary of SMSB stand simultaneously at the forefront and frontiers of contemporary soft matter science \cite{Yashima2016,Sang2022}. Recently, SMSB was observed in liquid crystals (LCs), where supramolecular chiral and polar structures are assembled from achiral, bent-shaped (curved) molecules \cite{Tschierske2018}. Bent-shaped molecules (BSMs) tend to form, i.a., smectic and nematic phases, whereas the latter have recently attracted a resurgence of worldwide interest due to novel findings \cite{Allesandro2017,Dozov2020}. One of them is the discovery of twist-bend nematic $(\text{N}_\text{TB})$ phase \cite{Cestari2011,Borshch2013,Chen2013} (Fig. \ref{fig:1}), which constitutes the first example of SMSB in a liquid state with no support from long-range spatial ordering. The characteristic feature of $\text{N}_\text{TB}$ is the emergence of heliconical (1D modulated) structure of nanoscale pitch with a ground-state exhibiting a degenerate sign of chirality (ambidextrous chirality), which means that each helix handedness is equiprobable. Further advances in the studies on $\text{N}_\text{TB}$ subsequently sprouted into a broad sub-field of liquid crystal research dedicated to novel nematic phases of non-standard symmetries \cite{Meyer2020,Rico2020,Mandle2021}.

It is noteworthy to point out that remarkable features of nematic phases formed by BSMs are not limited solely to the phenomenon of SMSB. The list is vast but just to name a few \cite{BSLC2017,RevModPhys.90.045004}: negative bend-splay elastic anisotropy (i.e. $K_{33}-K_{11} < 0$), giant flexoelectricity, high Kerr constants, and large rotational viscosity. The BSMs are also said to be the perfect candidates for optically biaxial fluid, namely biaxial nematic $(\text{N}_{B})$, predicted by Freiser in 1970 \cite{Freiser1970}. It is so because the introduction of the bend deformation in the molecular structure naturally specifies a secondary direction for orientational ordering, which is perpendicular to the long molecular axis \cite{Manoj2011,Lehmann2015}. 

All this makes LCs formed by BSMs, \emph{per se}, a riveting field to explore both from experimental and theoretical points of view \cite{Lubensky2002,Mettout2005,Ros2005,Mettout2007,RevModPhys.90.045004,Tomczyk2016,Tomczyk2020,Longa2020}. In particular, substantial efforts were undertaken to decipher how the molecule's structure influences the emergence of individual mesophases \cite{Reddy2006,Eremin2013,Lehmann2015}. One of valuable tools in such studies are computer simulations, such as Monte Carlo sampling (MC) and Molecular Dynamics (MD), which have a long-standing tradition in the investigation of structural properties of LC phases using a diverse range of models ranging from simplified lattice systems to fully atomistic ones \cite{Care_2005,Wilson2005,Allen2019}. Generally, the anisotropic shape of BSMs can be approximated in the simulations either by V-like or C-like bent objects (reflecting the $C_{2v}$ symmetry) composed of i.a., needles\cite{Tavarone2015,Karbowniczek2017,Ramirez2021}, spheres\cite{Janling2001,Dewar2004,Gregorio2016,GrecoFerrariniPRL}, or spherocylinders\cite{Camp1999,Lansac2003,Chiappini2019,Chiappini2021}. It has been shown that those systems, each one separately, produce a wealth of interesting nematic and smectic phases. However, especially appealing are the systems where the supramolecular chirality emerges out of pure entropy-driven interactions.

Herein, we have extended the work of Greco and Ferrarini \cite{GrecoFerrariniPRL} where it was shown that for a model system of BSMs (C-shape-like), composed of eleven tangent spheres -- see Fig.~\ref{fig:molecule}, it is possible to observe the stabilization of $\text{N}_\text{TB}$ arising from the packing entropy in MD simulations. We aim to provide an in-depth analysis of this system by altering the BSMs' curvature and packing densities in order to obtain the full phase diagram for the liquid state\footnote{The study of regions, where crystal structures emerge is outside of the article's scope and stands as the starting point for further research.}. To achieve it we performed both MC and MD simulations.      

The article is organized as follows. In Sec.~2 the system under study is described together with the details concerning the MC and MD simulations used to investigate its properties. The results of the simulations are given in Sec.~3 where they are compared with the current literature. In Sec.~4 we provide conclusions drawn from the acquired data.

\section{Model and simulation details}

\subsection{The model system}

\begin{figure}[ht]
    \centering
    \includegraphics[width=0.6\columnwidth]{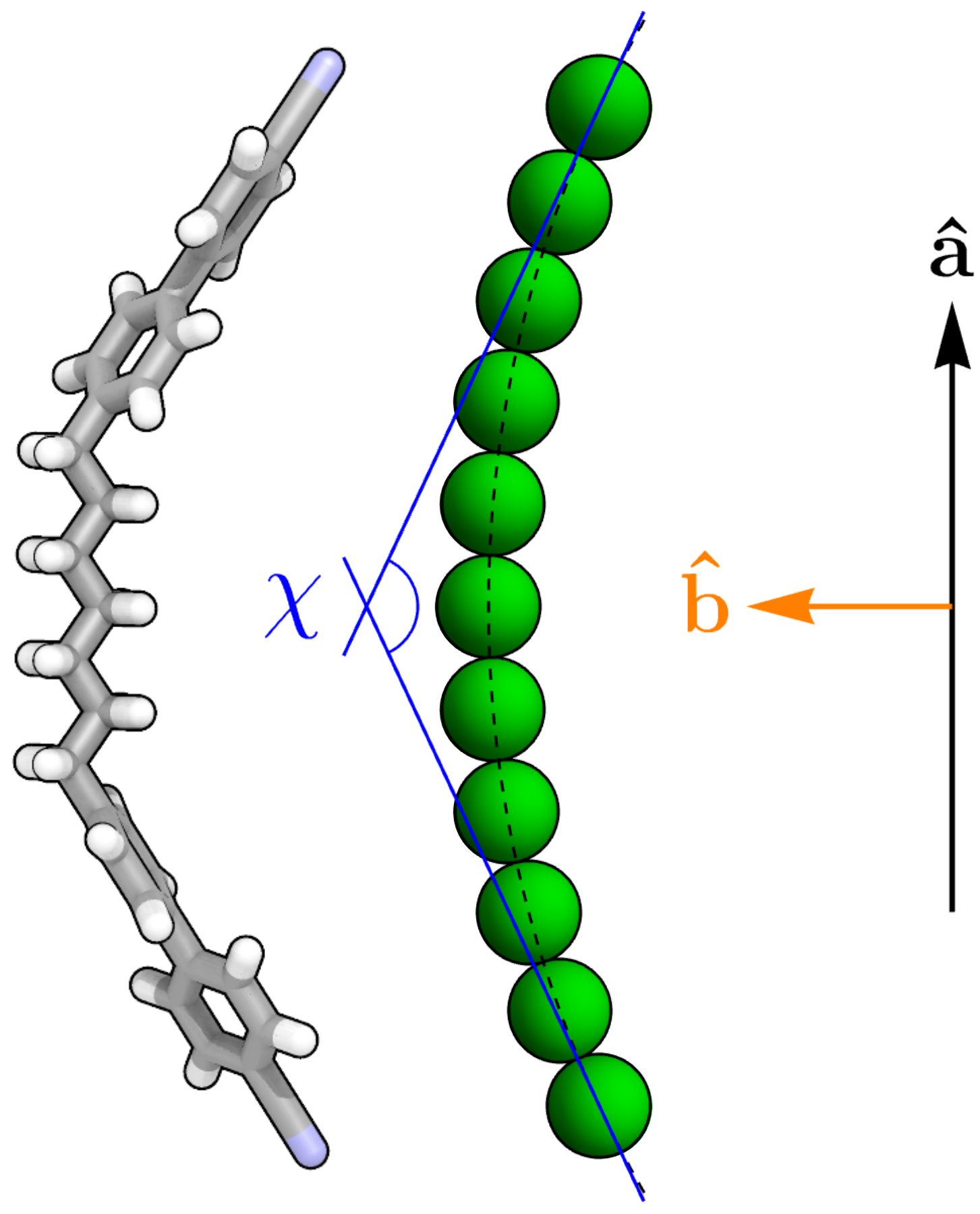}
    \caption{
    Left: Molecular structure of $\text{N}_\text{TB}$ forming mesogen CB7CB \cite{Cestari2011} rendered via QuteMol \cite{QuteMol}. Right: Model bent-shaped molecule with $C_{2v}$ symmetry composed of eleven tangent spheres whose centers are equidistantly distributed on the arc. The bend angle $\chi$ is defined as the angle formed by two lines, tangent to the arc at the centers of end-point spheres. Unit vectors $\mathbf{\hat{a}}$ and $\mathbf{\hat{b}}$ define the molecular frame, where $\mathbf{\hat{a}}$ equates with molecular long axis and $\mathbf{\hat{b}}$ is parallel to the molecule's two-fold $(C_2)$ symmetry axis.}
    \label{fig:molecule}
\end{figure}

BSMs were modeled by eleven tangent spheres of diameter $\sigma$ placed on a circular arc with a variable bend angle (see Fig.~\ref{fig:molecule}). For $\chi = 180^\circ$, the molecule reduces to a linear chain of adjoint spheres \cite{}, while for $\chi = 0^\circ$ it forms a semi-circle. To study lyotropic phase transitions in a system composed of such molecules, we examined two types of excluded-volume interaction. The first one is hard-core interaction \cite{Mederos2014}, while the second is Weeks-Chandler-Anderson (WCA) \cite{WCA1971,WCA1983} soft repulsion of each pair of spheres, defined as
\begin{equation}\label{eq:wca}
    U(r_{mn}) =
    \begin{cases}
        4\epsilon \qty[\qty(\frac{\sigma}{r_{mn}})^{12} - \qty(\frac{\sigma}{r_{mn}})^6] + \epsilon & \text{if $r_{mn} < 2^{1/6}\sigma$}\\
        0 & \text{if $r_{mn} \geq 2^{1/6}\sigma$},
    \end{cases}
\end{equation}
where $r_{mn}$ is the distance between the interaction centers and $\epsilon$ is the energy scale. The free parameters of the model $\sigma$, $\epsilon$ can be conveniently incorporated into the reduced pressure $P^* = P\sigma^3/\epsilon$ and reduced temperature $T^* = k_B T / \epsilon$. For hard-core repulsion, the equation of state depends only a single parameter -- the $P^*/T^* = P\sigma^3 / (k_B T)$ ratio \cite{Frenkel1987,Mederos2014}. For WCA interaction the system is sensitive to $P^*$ changes, while the dependence on $T^*$ is weak~\cite{Heyes2006}. The emergent phases were studied as a function of bending angle $\chi$ and packing density $\eta$. The latter one is a natural parameter for lyotropic systems and enables us to compare the two interaction models \cite{Heyes2006}. For both of them, the main contribution to Helmholtz free energy is the packing entropy: $F \approx -TS$ (the formula is exact for hard-core repulsion), so the phases with the highest entropy minimize $F$ and thus are promoted.

\subsection{Simulation methods}

The hard-core interactions were simulated using Monte Carlo sampling in $NPT$ ensemble~\cite{Wood1968}. The system was built of $N = 3000$ molecules. A single MC cycle consisted of $N$ molecule moves and a volume move. To perform a molecule move, a single molecule was sampled at random. Then, the molecule was translated by a random vector and rotated by a random angle around a random axis. The move was accepted provided it did not introduce an intersection. For volume moves all three lengths of an orthorhombic simulation box with periodic boundary conditions and the positions of molecules were scaled by three independent logarithmically distributed factors. Moves introducing overlaps were rejected and the rest of them was accepted according to Metropolis criterion. The extents of random moves were selected such that the acceptance ratio was around 15\%. The initial configuration was a diluted antiferroelectric sc crystal, which was then compressed or expanded to a target density. The equilibration was very sluggish due to concavities of molecules and required up to $2 \cross 10^8$ of full MC cycles. The production runs to sample ensemble averages consisted of $3 \times 10^7$ cycles. To accelerate the simulations, we parallelized them using domain division~\cite{Anderson2016}.

The MD simulations of $N=2025$ molecules, interacting \emph{via} the WCA potential in the $NPT$  ensemble at various values of the pressure $P^\ast$ in the interval $0.2\leq P^\ast \leq 2.0$, were performed using the open-source software LAMMPS \cite{LAMMPS}. All relevant simulation aspects were set up according to the framework provided in \cite{supplemental}. Equilibration and production runs were conducted for $10^7$ timesteps each.

\subsection{Order parameters}
Distinctive features of a particular phase can be characterized by appropriate order parameters and a corresponding degree of ordering. In nematic and smectic phases, the molecules' long axes $\vu{a}$ tend to orient, on average, along a preferred direction called director $\vu{n}$. Such kind of alignment can be quantified by the second-rank order parameter~\cite{Pasini1999, Allen2017}
\begin{equation}
    \expval{P_2} = \frac{1}{N} \expval{ \sum_{i=1}^{N} P_2(\vu{a}_i \vdot \vu{n}) },
\end{equation}
where the sum is over all $N$ molecules in the system, $P_2$ is the Legendre polynomial of degree $2$, $\vu{a}_i$ is a molecular axis vector of $i$-th molecule (cf. Fig.~\ref{fig:molecule}) and $\expval{\dots}$ denotes the time average over uncorrelated system snapshots. If $\expval{P_2} = 0$ orientations are isotropic, $\expval{P_2} = 1$ is a perfect order, and for $\expval{P_2} = -1/2$ all molecules are perpendicular to $\vu{n}$. Alternatively, $\expval{P_2}$ can be computed by diagonalising the second rank $\vb{Q}$-tensor \cite{deGennesBook}
\begin{equation}
     Q_{\alpha\beta} = \frac{1}{N} \sum_{i=1}^{N} \frac{3}{2}\qty(\hat{a}_{i,\alpha} \hat{a}_{i,\beta} - \frac{1}{3}\delta_{\alpha\beta}),
\end{equation}
where $\hat{a}_{i,\alpha}$ and $\hat{a}_{i,\beta}$ are the Cartesian components of $\mathbf{\hat{a}}_i$ and $\delta_{\alpha\beta}$ is the Kronecker delta. The nematic order parameter $P_2$ (for a single snapshot) is associated with the largest-modulus of a non-degenerate 
eigenvalue of $\vb{Q}$ and $\vu{n}$ is the corresponding eigenvector. In this formulation, zero-energy diffusive motion of $\vu{n}$ does not have effect on $\expval{P_2}$ value.

The density modulation is measured by smectic order parameter~\cite{Pasini1999,Allen2017}
\begin{equation}
    \expval{\tau} = \frac{1}{\rho} \expval{\abs{ \sum_{i=1}^{N} \exp(i \vb{k} \cdot \vb{r}_i)}},
\end{equation}
where $\rho$ is number density, $\vb{k}$ is a wavevector of density undulation and $\vb{r}_i$ are molecules' centers of mass. Typically, for non-tilted smectic phases $\vb{k} \parallel \vu{n}$. For $\expval{\tau} = 0$ there is no density modulation, whereas for $\expval{\tau} = 1$ the molecules form perfect layers with a spacing equal to $2\pi/\abs{\vb{k}}$.

Total polarization can be computed as a sum of molecule polarization vectors
\begin{equation}
    \expval{\vb{M}} = \expval{ \sum_{i=1}^{N} \vu{b}_i }.
\end{equation}
As it will be seen from the results, the non-zero polarization can be present when one restricts the summation to a plane perpendicular to $\vu{n}$, however, at the same time, it can vary in the direction of $\vu{n}$ and give zero net polarization when summed over the whole system. To account for such variations, one can calculate the norm $M(C_j) = \norm{\vb{M}(C_j \pm \Delta C/2)}$ for molecules in a narrow shell between two planes $\vb{r} \vdot \vu{n} = C_j \pm \Delta C/2$ and sum them over the whole system
\begin{equation}
    \expval{m} = \frac{1}{N_b} \expval{ \sum_{j=1}^{N_b} M(C_j) }.
\end{equation}
Here, $N_b = L/\Delta C$ is the number of shells and $L$ is the system length in the direction of $\vu{n}$.

\section{Results}

\subsection{Phase overview}

\begin{figure}[ht]
    \centering
    \includegraphics[width=0.8\linewidth]{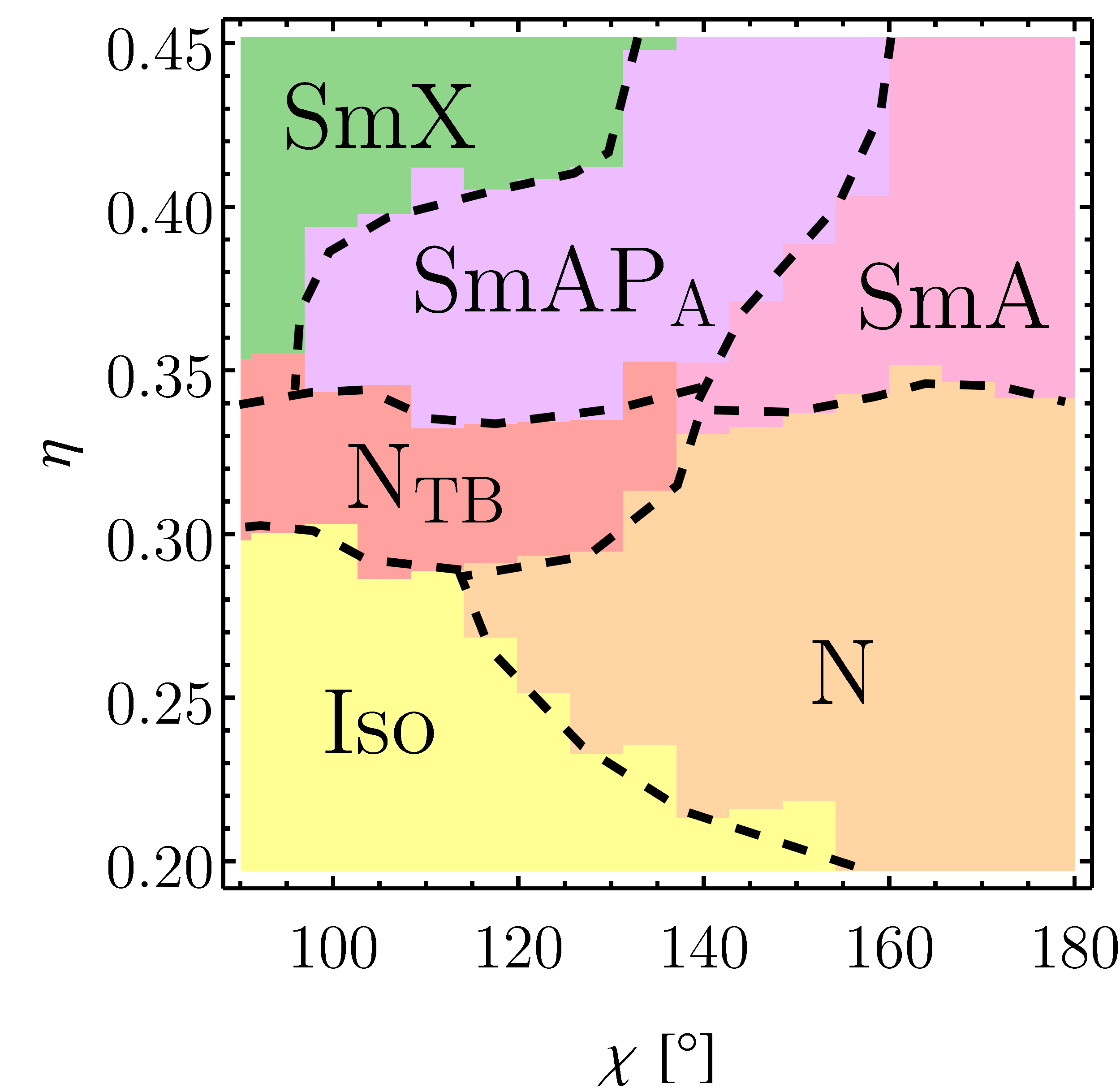}
    \caption{Phase diagram for hard-core interactions in $(\chi,\eta)$ parameters space. The following phases are present: isotropic liquid (Iso), nematic (N), twist-bend nematic ($\text{N}_\text{TB}$), non-polar smectic A (SmA), antiferroelectric smectic A ($\text{SmAP}_\text{A}$) and unidentified polar smectic (SmX). Black dashed lines are arbitrarily drawn to visually separate the phases.}
    \label{fig:pd}
\end{figure}

\begin{figure*}[ht]
    \centering
    \includegraphics[height=0.27\linewidth]{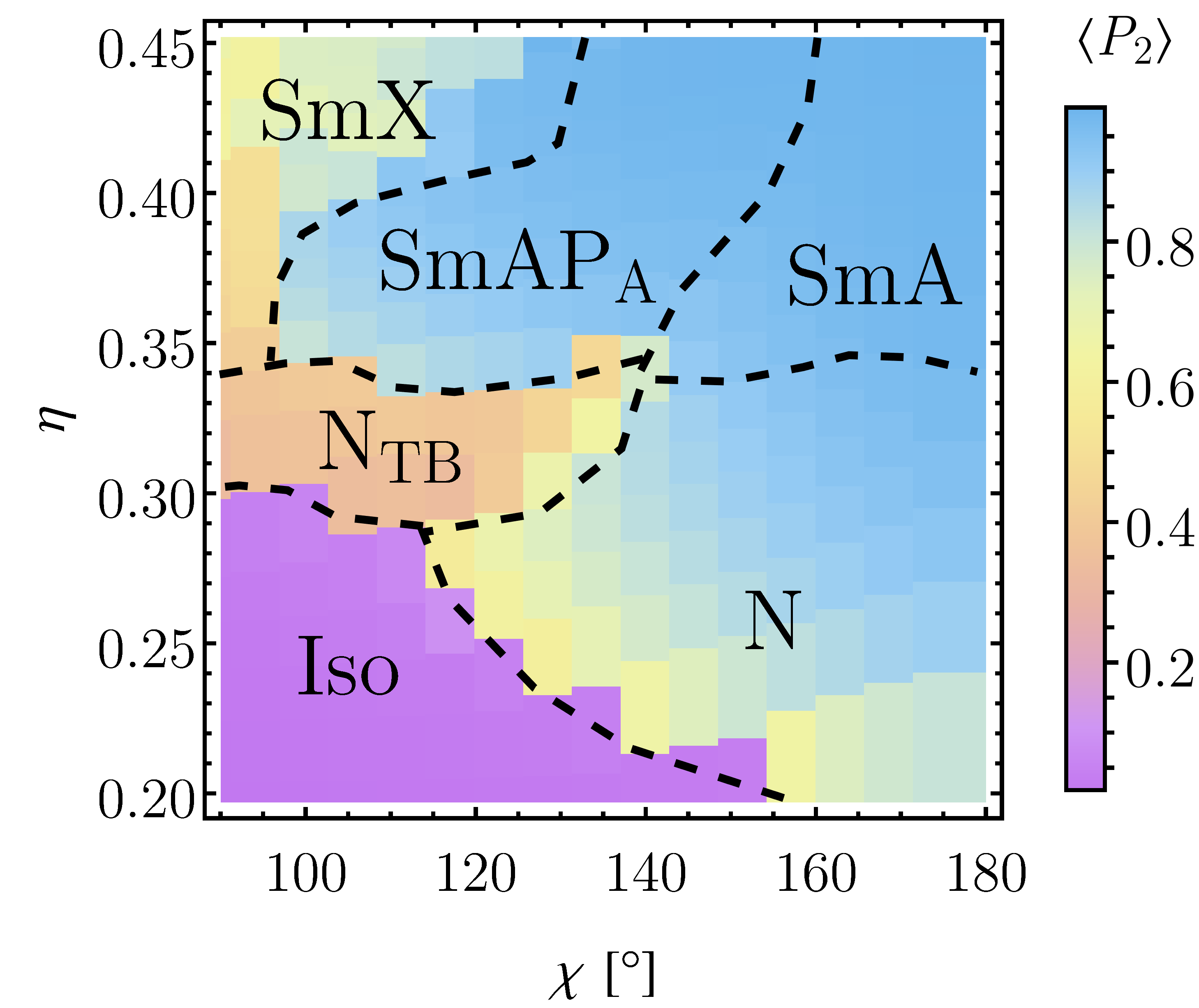}
    \includegraphics[height=0.27\linewidth]{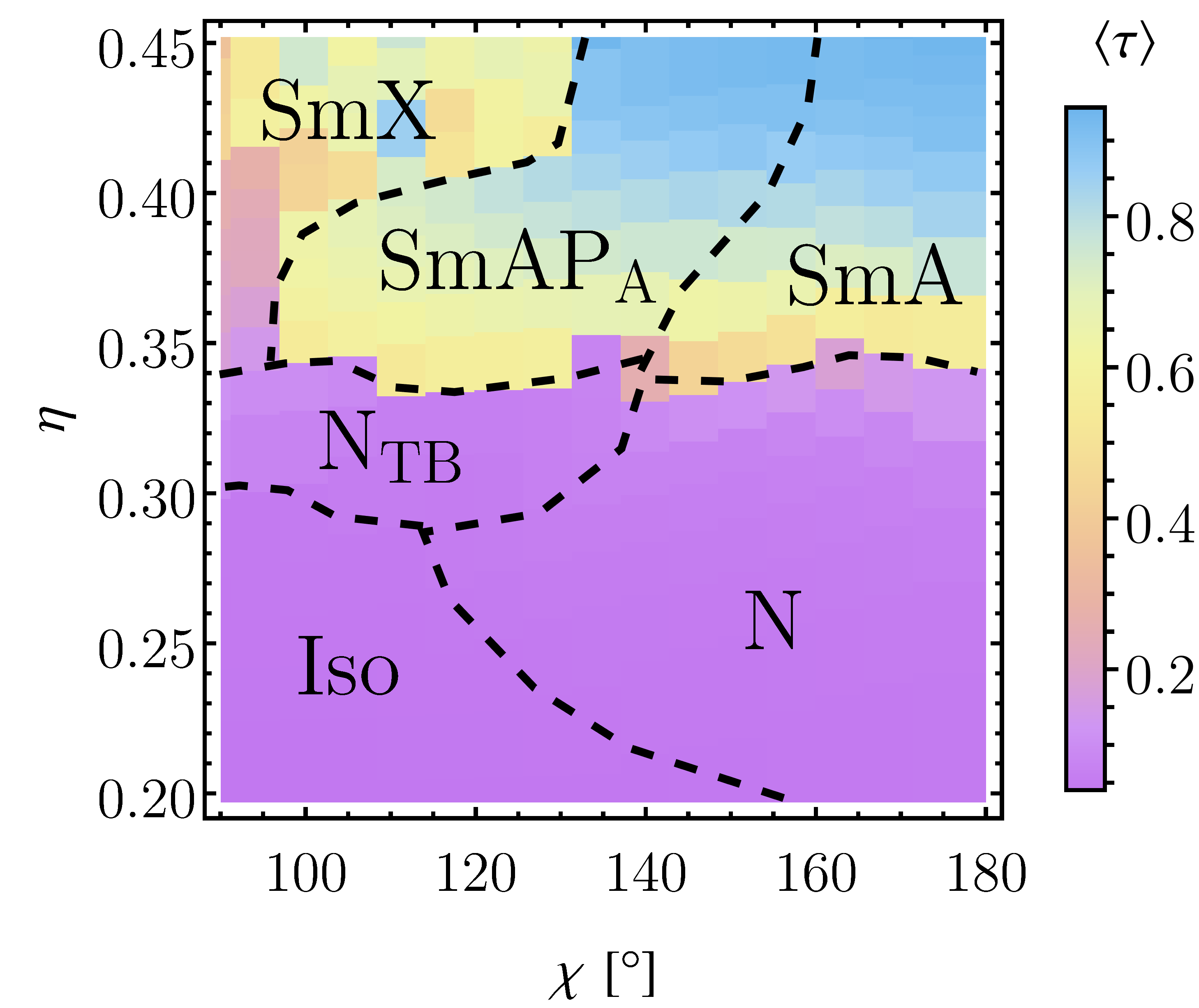}
    \includegraphics[height=0.27\linewidth]{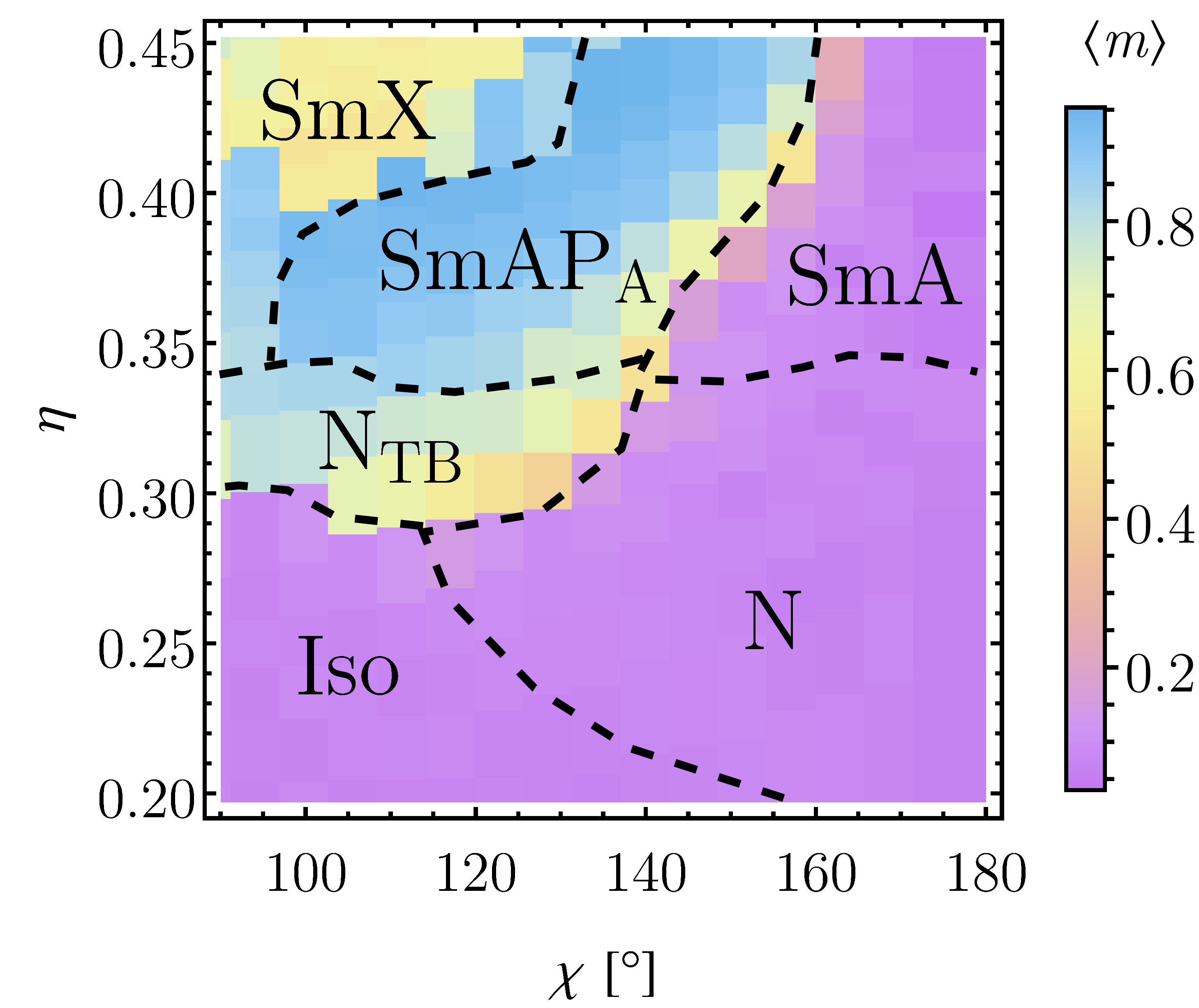}
    \caption{Ensemble averages of order parameters as a function of $\chi$ and $\eta$. (a) Nematic order parameter $\expval{P_2}$, (b) smectic order parameter $\expval{\tau}$, (c) layer polarization $\expval{m}$.}
    \label{fig:order_params}
\end{figure*}

\begin{figure*}[p]
    \centering
    \includegraphics[height=0.8\paperheight]{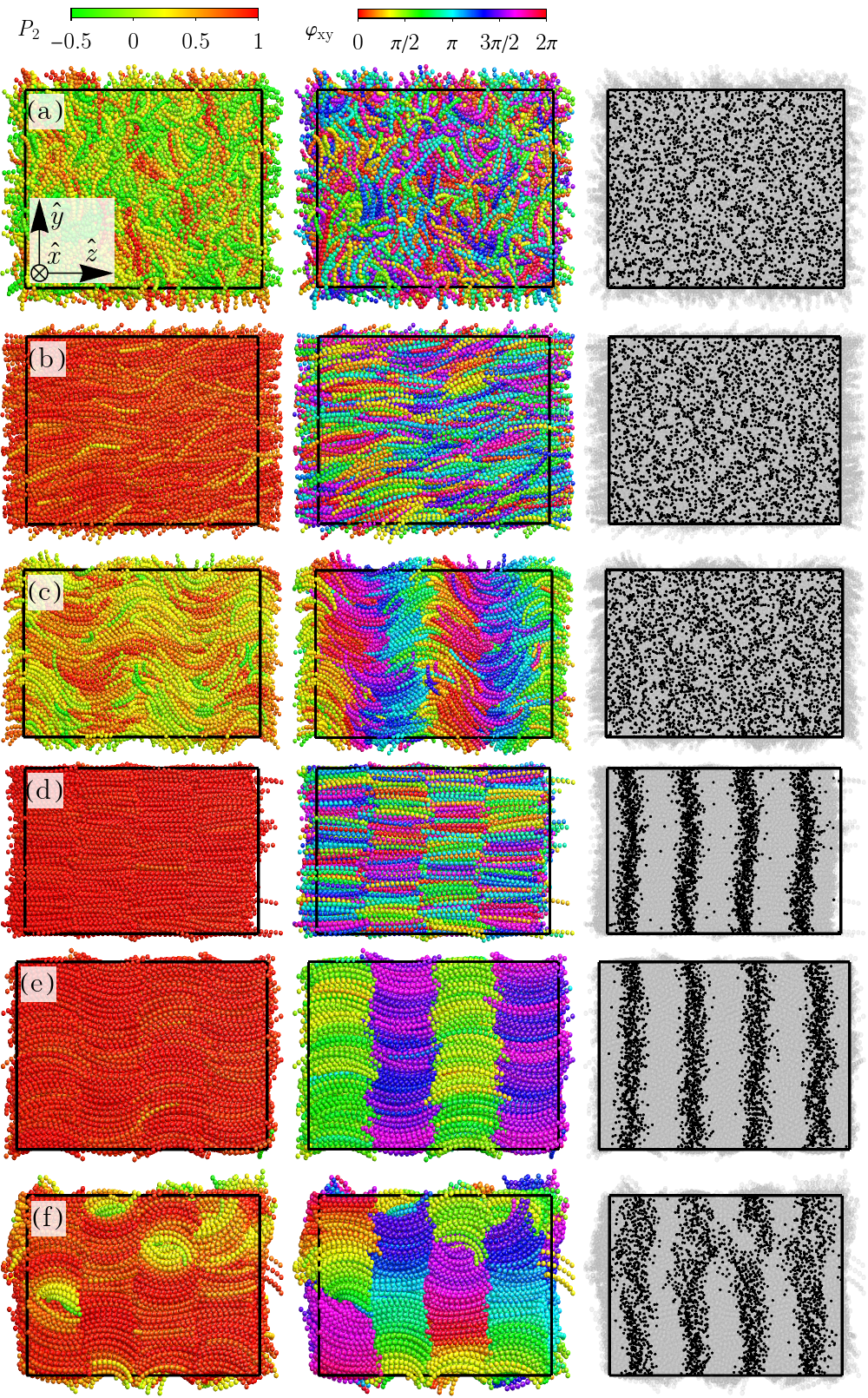}
    \caption{System snapshots for all identified phases. Each row presents one phase, from the top:
    (a) Iso [$(\chi,\eta)=(95^\circ,0.27)$],
    (b) N [$(\chi,\eta)=(140^\circ,0.31)$],
    (c) $\text{N}_\text{TB}$ [$(\chi,\eta)=(100^\circ,0.32)$],
    (d) SmA [$(\chi,\eta)=(160^\circ,0.39)$],
    (e) $\text{SmAP}_\text{A}$ [$(\chi,\eta)=(130^\circ,0.39)$], (f) SmX [$(\chi,\eta)=(100^\circ,0.41)$]. In the first column molecules are color-coded according to $P_2$ value, in the second one -- according to molecule polarization angle $\phi_{xy}$ between $x$ axis and the projection of $\vb{b}$ molecule polarization vector onto $xy$ plane. Finally, in the last column, black dots mark molecule mass centers.}

    \label{fig:packings}
\end{figure*}

The system exhibits a variety of liquid phases: isotropic liquid (Iso), nematic (N), twist-bend nematic $(\text{N}_\text{TB})$, non-polar smectic A (SmA), antiferroelectric smectic A ($\text{SmAP}_\text{A}$) and an unidentified polar smectic (SmX). The phase diagram for hard-core interactions is shown in Fig.~\ref{fig:pd}. WCA soft repulsion recreates all the phases with nearly identical phase borders. All order parameters described in the previous section are gathered in Fig.~\ref{fig:order_params} as a function of the bend angle $\chi$ and the packing fraction $\eta$. Finally, system snapshots for all identified phases are depicted in Fig.~\ref{fig:packings}.

For each value of $\chi$ and sufficiently low $\eta$, the molecules form ordinary isotropic liquid as shown by low values of all order parameters indicating lack of any long-range correlations [see Fig.~\ref{fig:packings}(a)]. Around $\eta = 0.45$ the system crystallizes into several types of hexagonal close-packed crystals, which is beyond the scope of this work. The boundary of the Iso phase moves upwards in $\eta$ for a decreasing opening angle $\chi$. It can be explained by the fact, that a lower bend angle results in less prolate molecules, for which the entropic gain from orientational ordering is lower. 

\subsection{Nematic phases}

The ordinary nematic phase N can be observed for $\chi > 110^\circ$, as indicated by a sharp jump of $\expval{P_2}$ value from near zero to over 0.5 on the phase boundary [see also Fig.~\ref{fig:packings}(b)]. A growing $\eta$ results in an increase of $\expval{P_2}$ to almost $0.9$ for $\chi \approx 180^\circ$, which is significantly higher than a typical range $[0.3, 0.7]$ for nematics observed in experiments \cite{Chandrasekhar1980}. In this range of parameters $(\chi, \eta)$, no density modulation is observed. Moreover, two smallest eigenvalues of $\expval{\vb{Q}}$ are comparable, meaning that the nematic is uniaxial, which is further supported by a vanishing net polarization $\expval{\vb{M}} \approx \vb{0}$ (result not shown). Layer polarization $\expval{m}$ is also near zero, so the phase is homogeneous.

The situation changes for a phase forming over the Iso phase for $\chi < 110^\circ$ and over the N phase in the range $\chi \in [110^\circ, 140^\circ]$, for $\eta$ values between 0.29 and 0.35 [see Fig.~\ref{fig:packings}(c)]. No density modulation suggests that is is a nematic, however with $\expval{P_2} \in [0.3, 0.5]$, lower than for the ordinary N phase in the system. The net polarization is zero, while $\expval{m}$ lies in the range $[0.4, 0.8]$, indicating high polar ordering in planes perpendicular to $\vu{n}$. A closer inspection reveals that this is twist-bend nematic phase $\text{N}_\text{TB}$, observed in many studies of bent-core molecules~\cite{Memmer2002, Shamid2013statistical, GrecoFerrariniPRL, Chiappini2021} (see also Fig.~\ref{fig:1}). In this phase, both $\vu{n}$ and polarization vary in space in a heliconical pattern. Assuming the global director is aligned with the $z$ axis, the director field may be parameterized as
\begin{equation}\label{eq:n_tb}
    \vu{n}(x, y, z) = \mqty(\sin(\theta) \cos(2\pi z/p) \\
                            \sin(\theta) \sin(2\pi z/p) \\
                            \cos(\theta)).
\end{equation}
The spatial dependence of $\vb{M}$ is similar, with $\vb{M} \perp \vb{n}$. $\text{N}_\text{TB}$ phase has two parameters -- the conical angle $\theta$, which is the tilt of the director from the $z$ axis, and a pitch $p$ defining the periodicity of the structure. The parameters vary in phase space -- $p \in [25\sigma, 45\sigma]$ and $\theta \in [25^\circ, 35^\circ]$. With increasing packing fraction $\eta$ or decreasing bend angle $\chi$ the period $p$ shortens and $\theta$ becomes more obtuse. $\text{N}_\text{TB}$ can be predicted using phenomenological lowest-order Oseen-Zocher-Frank free energy \cite{Oseen1933,Zocher1933,Frank1958}
\begin{equation}
    F_{\text{OZF}} = \frac{1}{2}K_{11} \qty[\vu{n}(\div{\vu{n}})]^2 + \frac{1}{2}K_{22} \qty[\vu{n}\vdot(\curl{\vu{n}})]^2 + \frac{1}{2}K_{33} \qty[\vu{n} \cp (\curl{\vu{n}})]^2,
\end{equation}
where subsequent terms represent, respectively, splay, twist and bend deformations of director field and $K_{ii}\;(i=1,2,3)$ are Frank elastic constants. All three Frank elastic constants have to obey Ericksen inequalities \cite{Ericksen1966}, i.e.: $K_{ii} \geq 0\;(i=1,2,3)$, otherwise perfect alignment (homogeneous nematic) would not correspond to a state of minimum
energy. However, if one were to consider scenario where $K_{33}<0$, then it would occur that bend deformation is energetically favorable and reference state, i.e. homogeneous nematic, becomes unstable \cite{Dozov_2001}. As there does not exist a structure with constant bend without topological defects, is has to be combined with at least one of: splay or twist deformations. If $K_{11} > 2K_{22}$, twist is preferred over splay~\cite{Dozov_2001}. As pointed by Shamid \emph{et al.} \cite{Shamid2013statistical}, the effective $K_{33}$ constant can become negative below a certain critical temperature, when the coupling of polarization with bend deformation is introduced, supporting the assumption that $\text{N}_\text{TB}$ is inherently polar. The phase can be also understood via geometrical reasoning \cite{Tomczyk2016,Longa2016,Tomczyk2020,Longa2020}. For curved spherocylinders, one can predict a correct relation between molecule bend angle, pitch $p$ and conical angle $\theta$ matching the spherocylinder curvature with an integral curve of a director field $\vu{n}$~\cite{Chiappini2021}.

\subsection{Smectic phases}

For $\eta > 0.34$, smectic phases are formed. The N phase is adjacent to an ordinary SmA phase -- $\expval{\tau}$ is over $0.5$ and strata are clearly visible [see Fig.~\ref{fig:packings}(d)]. Further compression narrows the spread of molecule mass centers in the direction normal to the layer and $\tau$ approaches unity for a high $\eta$. Typically, $\expval{P_2}$ value experiences a jump on the N-SmA boundary from around 0.4-0.6 to over 0.8 \cite{Memmer2002}, however in some systems it is small or absent \cite{Lansac2003}. Here, no sudden change of $\expval{P_2}$ is observed. Molecule polarizations are chaotic within the layers, as indicated by a low value of $\expval{m}$ and they are orthogonal to $\vu{n}$. 

\begin{figure*}[htb]
    \centering
    \includegraphics[width=0.8\linewidth]{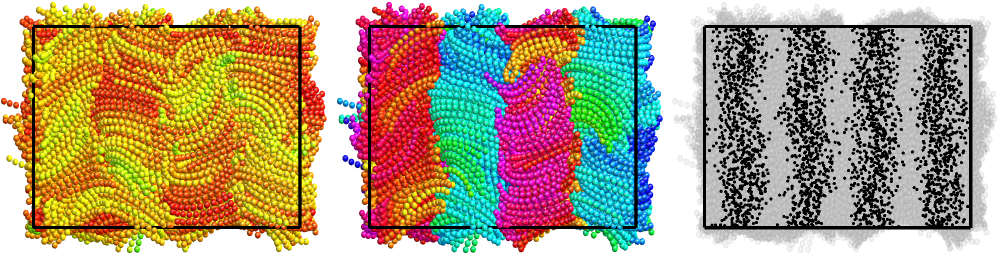}
    \caption{System snapshot for $\text{SmC}_\text{A}\text{P}_\text{A}$ phase [$(\chi,\eta)=(65^\circ,0.45)$] in analogy to Fig.~\ref{fig:packings}.
    }
    \label{fig:SmCAPA}
\end{figure*}

A distinct smectic phase is formed over $\text{N}_\text{TB}$ phase for $\chi > 100^\circ$ and over SmA for $\chi \in [140^\circ, 160^\circ]$ [see Fig.~\ref{fig:packings}(e)]. A high value of $\expval{m}$ indicates that it is a polar phase and a closer inspection reveals that adjacent layers have an opposite (antiferroelectric) polarization, allowing us to classify the phase as $\text{SmAP}_\text{A}$. The onset of polar order may be attributed to the entropic gain of translational motion within a layer, however, it does not explain the antiferroelectric arrangement of layer polarization. Lansac et al.~\cite{Lansac2003} attributed such a configuration to the more compatible vibrational motion of two molecules from adjacent layers when they have antiparallel directions compared to parallel ones, resulting in once again in a higher entropy of the former system. $\text{SmAP}_\text{A}$-SmA boundary moves to higher densities for less bent molecules -- lower curvatures discourage polar order, so a stronger compression is required to achieve it. In particular, for $\chi > 160^\circ$, a crystallization occurs before the formation of $\text{SmAP}_\text{A}$\footnote{Notably, the crystal is polar for $\chi \in [160^\circ, 170^\circ]$ and non-polar above $\chi = 170^\circ$.}. Note that $\text{N}_\text{TB}$ always forms a polar smectic when compressed. It can be expected, since $\text{N}_\text{TB}$ is already polar. A further insight into the connection between $\text{SmAP}_\text{A}$ and $\text{N}_\text{TB}$ can be given by the analysis of $\vu{n}(x,y,z)$ spatial dependence. The numerical data is well described by a director field
\begin{equation}\label{eq:sm_sb}
    \vu{n}(x, y, z) = \mqty(\sin(\theta) \cos(2\pi z/p) \\
                            0 \\
                            \sqrt{1 - \sin^2(\theta) \cos^2(2\pi z/p)}),
\end{equation}
thus the emergent $\text{SmAP}_\text{A}$ phase can also be described as splay-bend smectic ($\text{Sm}_\text{SB}$). There, the director is constant in $xy$ plane and oscillates within $xz$ plane around $\vu{z}$ direction, coinciding with the direction $\vb{k}$ of density modulation. Although $\vu{n}$ leans from $\vb{k}$ within a layer, which would suggest type C smectic, averaging over the whole layer gives $\vu{n} \parallel \vb{k}$ indicating that this is SmA, as previously stated. The director fields \eqref{eq:n_tb}, \eqref{eq:sm_sb} are special cases of a more general director field modulation with all of splay, twist, and bend deformations
\begin{equation}\label{eq:sm_tsb}
    \vu{n}(x, y, z) = \mqty(\sin(\theta_a) \cos(2\pi z/p) \\
                            \sin(\theta_b) \sin(2\pi z/p) \\
                            \sqrt{1 - \sin^2(\theta_a) \cos^2(2\pi z/p) - \sin^2(\theta_b) \sin^2(2\pi z/p)}),
\end{equation}
where the angles $\theta_a$ and $\theta_b$ may be different. Fixing $\theta_a = \theta$ and continuously changing $\theta_b$ from 0 to $\theta$ interpolates between director fields for $\text{N}_\text{TB}$ and $\text{Sm}_\text{SB}$. 
This intermediate splay-twist-bent smectic $\text{Sm}_\text{STB}$ phase has been observed for bent spherocylinders~\cite{Chiappini2021}. In our model, however, it is absent.

Highly bent molecules form yet another polar smectic phase, as indicated by high $\expval{\tau}$ and $\expval{m}$ values, however of an unknown type, denoted as SmX [see Fig.~\ref{fig:packings}(f)]. Despite trying both compression and expansion from various initial configurations (dense crystal, diluted crystal, type C smectic, lower-density equilibrium system, and others) we were not able to obtain a structure without irregular domain walls. As it can be seen from Fig.~\ref{fig:packings}, the polarization within each stratus rotates with $y$ coordinate and topological defects are visible. The same effects are present for both hard and soft interactions. Increasing the system size, which allows a full rotation of the molecule polarization vector has not eliminated the defects and in neither case they form a regular pattern as in e.g. blue phases~\cite{Wright1989}. However, the blue-phase-like structures cannot be completely disregarded since the lattice constant of the defects can be of the order of tens or even hundreds of molecule lengths, and the system size allowing such long periods is far beyond the limit of our computational resources. Another candidate is anticlinic, antiferroelectric type C smectic ($\text{SmC}_\text{A}\text{P}_\text{A}$) \cite{Gimeno2014}. There, both the director tilt and polarization alternate between layers. Remarkably, we have observed its formation for a very acute bend angle $\chi < 70^\circ$ and a high density $\eta > 0.4$ (see Fig.~\ref{fig:SmCAPA}). 

BSMs with highly acute bend angle are still scarce and not fully understood \cite{Gimeno2014,BSLC2017a}. Additionally, chemical data shows that increasing the degree of bend diminishes the LC phase stability to the point where the molecular bend is so acute that the compound becomes non-mesogenic \cite{Hird2005}. Therefore, $\chi < 90^\circ$ falls outside of the scope of this work. Nevertheless, further investigation of the unidentified SmX phase can be a potential goal for future studies.

\section{Conclusions}
We have systematically studied the phase diagram of the liquids formed by bent-shaped molecules composed of tangential spheres with centers placed on an arc. The molecules were characterized according to adjustable bend angle $\chi$, and the system of them was considered under different packing densities $\eta$. In $(\chi, \eta)$ parameters space, we have identified isotropic, nematic, twist-bend nematic, smectic A, and antiferroelectric smectic A phases. For the highest density and the largest $\chi$ we have observed a smectic phase, denoted as SmX, forming polar layers with in-layer rotating polarization. We were, however, not able to avoid defects and domain walls. For unrealistically large bend, SmX phase transformed to anticlinic, antiferroelectric type C smectic, where both the director tilt and polarization alternate between layers. In comparison to a previous study for bent spherocylinders~\cite{Chiappini2021}, we have not observed splay-twist-bend smectic. The obtained results were the same for both: MC with hard-core interactions and MD using short-range WCA repulsion.

\section*{Author Contributions}
P.K.: conceptualization, data curation, formal analysis, funding acquisition, investigation, project administration, software, visualization, writing. W.T.: conceptualization, data curation, formal analysis, investigation, validation, writing.
M.C.: conceptualization, writing. 

\section*{Conflicts of interest}
There are no conflicts of interest to declare.

\section*{Acknowledgements}
P.K. and M.C. acknowledge the support of Ministry of Science and Higher Education (Poland) grant no. 0108/DIA/2020/49. Numerical simulations were carried out with the support of the Interdisciplinary Center for Mathematical and Computational Modeling (ICM) at the University of Warsaw under grant no. GB76-1.



\balance


\bibliography{References} 
\bibliographystyle{rsc} 

\end{document}